\documentclass[12pt]{iopart}
\usepackage{color}
\usepackage{cite}
\usepackage{braket}
\usepackage[pdftex]{graphicx}

\usepackage{iopams}  

\begin{document}

\title[]{Improved absolute frequency measurement of $^{171}$Yb at NMIJ with uncertainty below $2\times10^{-16}$}

\author{Takumi Kobayashi$^{1,*}$, Akiko Nishiyama$^{1}$, Kazumoto Hosaka$^{1}$, Daisuke Akamatsu$^{2}$, Akio Kawasaki$^{1}$, Masato Wada$^{1}$, Hajime Inaba$^{1}$, Takehiko Tanabe$^{1}$, Masami Yasuda$^{1}$}

\address{$^1$National Metrology Institute of Japan (NMIJ), National Institute of Advanced Industrial Science and Technology (AIST), 1-1-1 Umezono, Tsukuba, Ibaraki 305-8563, Japan\\
$^2$Department of Physics, Graduate School of Engineering Science, Yokohama National University, 79-5 Tokiwadai, Hodogaya-ku, Yokohama 240-8501, Japan\\}
\ead{takumi-kobayashi@aist.go.jp}
\vspace{10pt}
\begin{indented}
\item[\today]
\end{indented}

\begin{abstract}
We report improved absolute frequency measurement of the $^{1}$S$_{0}-^{3}$P$_{0}$ transition of $^{171}$Yb at National Metrology Institute of Japan (NMIJ) by comparing the $^{171}$Yb optical lattice clock NMIJ-Yb1 with 13 Cs primary frequency standards via International Atomic Time from August 2021 to May 2023. The measured absolute frequency is 518 295 836 590 863.62(10) Hz with a fractional uncertainty of $1.9\times10^{-16}$, in good agreement with the recommended frequency of $^{171}$Yb as a secondary representation of the second. This uncertainty is 2.6 times lower than our previous measurement uncertainty, and slightly lower than any uncertainties of the absolute frequency measurements of $^{171}$Yb that have so far been reported by other institutes. We also estimate correlation coefficients between our present and previous measurements, which is important for updating the recommended frequency. 
\end{abstract}

%
%
%
%
%

\section{Introduction}
Optical clocks have achieved fractional uncertainties at the $10^{-18}$ to $10^{-19}$ level \cite{Ushijima2015,Hunteman2016,McGrew2018,Brewer2019,Bothwell2019,Tofful2024,Li2024,Aeppli2024} and are considered as candidates for a redefinition of the second in the International System of Units (SI) \cite{Hong2016,Dimarcq2024}. The remarkable performances of optical clocks support new applications including studies of fundamental physics \cite{Wcislo2018,Takamoto2020,Kennedy2020,Lange2021,Kobayashi2022,Filzinger2023} and relativistic geodesy \cite{Takano2016,Grotti2018,Grotti2024}.

Towards the redefinition of the SI second, it is essential to ensure continuity with the current definition based on Cs, which is listed as one of mandatory criteria for the redefinition determined by the Consultative Committee for Time and Frequency (CCTF) \cite{Dimarcq2024}. To fulfill this criterion, at least three independent absolute frequency measurements of optical frequency transitions with uncertainties limited by Cs frequency standards ($<3\times10^{-16}$) are required. While measurements at this level have already been reported by several groups \cite{Grebing2016,Lodewyck2016,McGrew2019,Schwarz2020,Lange2021,Nemitz2021,Kim2021,Pizzocaro2019,Goti2023}, it is important to add new measurement results for confirming the consistency between independently developed clocks with different optical transitions \cite{Lodewyck2019}. Regarding the $^{1}$S$_{0}-^{3}$P$_{0}$ transition of the $^{171}$Yb optical lattice clock, absolute frequency measurements with uncertainties of $\lesssim3\times10^{-16}$ have been reported by National Institute of Standards and Technology (NIST) \cite{McGrew2019}, Istituto Nazionale di Ricerca Metrologica (INRIM) \cite{Pizzocaro2019,Clivati2022,Goti2023}, and Korea Research Institute of Standards and Science (KRISS) \cite{Kim2021}. At National Metrology Institute of Japan (NMIJ), we have recently reported two absolute frequency measurements of Yb using a satellite link to International Atomic Time (TAI) from 2019 to 2020 \cite{Kobayashi2020} and our local Cs fountain clock NMIJ-F2 from 2020 to 2021 \cite{Kobayashi2022}. The uncertainty of the first measurement using the TAI link was $5.0\times10^{-16}$, mostly limited by the systematic uncertainty of our Yb optical lattice clock NMIJ-Yb1 ($4\times10^{-16}$) \cite{Kobayashi2018}. The second measurement with NMIJ-F2 was carried out after improving the systematic uncertainty of NMIJ-Yb1 to $1\times10^{-16}$. The uncertainty of the second measurement was $5.3\times10^{-16}$ which in this time was mainly limited by the systematic uncertainty of NMIJ-F2 ($4.7\times10^{-16}$) \cite{Takamizawa2022}.

In this paper, we report a new absolute frequency measurement of Yb with the improved systematic uncertainty of NMIJ-Yb1 ($1\times10^{-16}$ \cite{Kobayashi2022}) using the link to TAI from 2021 to 2023. During this period, NMIJ-Yb1 is operated for calibrating the frequency of TAI as a secondary frequency standard (SFS). The absolute frequency of Yb is determined by comparing the frequency of TAI evaluated by NMIJ-Yb1 and those by 13 Cs primary frequency standards (PFSs). The uncertainty of the determined absolute frequency is $1.9\times10^{-16}$, which is slightly lower than the uncertainties of the previous absolute frequency measurements of Yb reported by NIST, INRIM, and KRISS.

\section{Method}
\subsection{Frequency measurement chain}
\begin{figure}[t]
\includegraphics[scale=0.5]{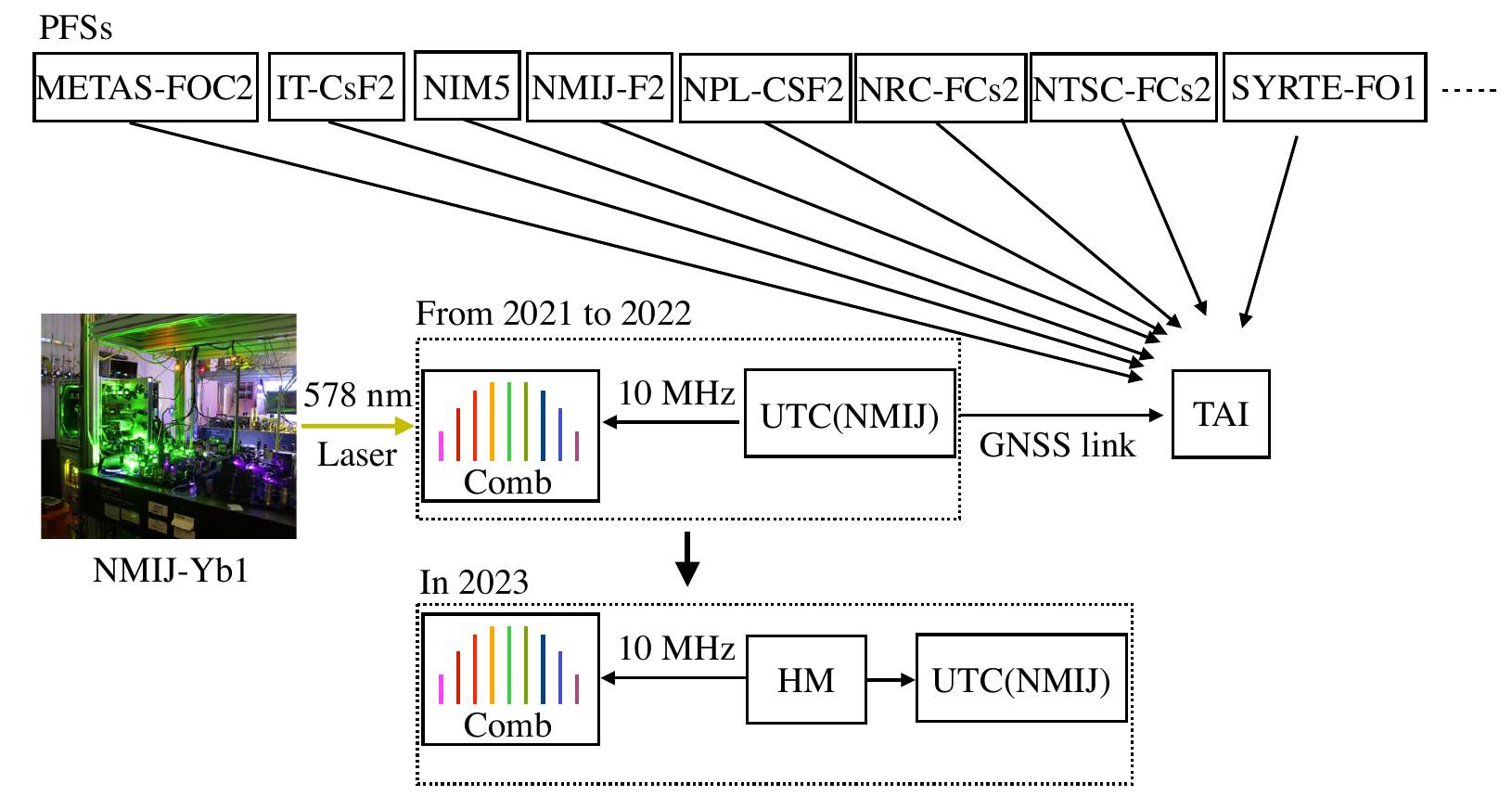}
\caption{Frequency measurement chain for the absolute frequency measurement of NMIJ-Yb1 referenced to PFSs via TAI.}
\label{experimentalsetup}
\end{figure}
Figure \ref{experimentalsetup} shows the frequency measurement chain for the absolute frequency measurement of NMIJ-Yb1 using the TAI link that was conducted from August 2021 to May 2023. The details of the experimental apparatus are described elsewhere \cite{Kobayashi2018,Kobayashi2020}. From 2021 to 2022, the frequency of NMIJ-Yb1 was compared with that of Coordinated Universal Time of NMIJ (UTC(NMIJ)) using an optical frequency comb \cite{Inaba2006}. From this comparison, the fractional frequency difference between NMIJ-Yb1, which is hereafter simply denoted by Yb in equations, and UTC(NMIJ) is calculated by
\begin{eqnarray}
y(\mathrm{Yb - UTC(NMIJ)})_{T_{\mathrm{Yb}}}&=&\frac{f^{\mathrm{a}}(\mathrm{Yb})}{f^{\mathrm{n}}(\mathrm{Yb})}-\frac{f^{\mathrm{a}}(\mathrm{UTC(NMIJ)})}{f^{\mathrm{n}}(\mathrm{UTC(NMIJ)})}\nonumber\\
&\sim&\frac{f^{\mathrm{a}}(\mathrm{Yb})/f^{\mathrm{a}}(\mathrm{UTC(NMIJ)})}{f^{\mathrm{n}}(\mathrm{Yb})/f^{\mathrm{n}}(\mathrm{UTC(NMIJ)})}-1,
\end{eqnarray}
where $y(\mathrm{A}-\mathrm{B})_{T}$ denotes the fractional frequency difference between A and B averaged over a period of $T$, and $f^{\mathrm{a(n)}}(\mathrm{A})$ the actual (nominal) frequency of A. The approximation is valid when $(f^{\mathrm{a}}(\mathrm{A})-f^{\mathrm{n}}(\mathrm{A}))\ll f^{\mathrm{n}}(\mathrm{A})$. $T_{\mathrm{Yb}}$ is the period of the operation of NMIJ-Yb1 which is not perfectly continuous. The nominal frequency is chosen as  $f^{\mathrm{n}}(\mathrm{UTC(NMIJ)})=10$ MHz and $f^{\mathrm{n}}(\mathrm{Yb})=f^{\mathrm{CIPM}}(\mathrm{Yb})=$ 518 295 836 590 863.63 Hz which is the CIPM (Comit\'e International des Poids et Mesures)
recommended frequency updated in 2021 \cite{Margolis2024}. UTC(NMIJ) is linked to TAI based on global navigation satellite systems (GNSSs), yielding the frequency difference between NMIJ-Yb1 and TAI by
\begin{eqnarray}
y(\mathrm{Yb-TAI})_{T_{\mathrm{link/Yb}}}&=&y(\mathrm{Yb - UTC(NMIJ)})_{T_{\mathrm{link/Yb}}}\nonumber\\
&&+y(\mathrm{UTC(NMIJ) - TAI})_{T_{\mathrm{link/Yb}}}.
\label{ybtaieq}
\end{eqnarray}
$y(\mathrm{UTC(NMIJ) - TAI})_{T_{\mathrm{link/Yb}}}$ is derived from time differences between UTC(NMIJ) and TAI which are provided in Circular T \cite{circulart}, a monthly report issued by Bureau International des Poids et Mesures (BIPM), by the relationship
\begin{equation}
y(\mathrm{UTC(NMIJ) - TAI})_{T_{\mathrm{link/Yb}}}=\frac{\Delta t_{\mathrm{f}} - \Delta t_{\mathrm{i}}}{T_{\mathrm{link/Yb}}}.
\label{phasefreqdiffeq}
\end{equation}
where $\Delta t_{\mathrm{i}}$ and $\Delta t_{\mathrm{f}}$ denote the initial and final time difference data for an interval $T_{\mathrm{link/Yb}}$, respectively.
In Circular T, the time differences are provided at 5-day intervals on Modified Julian Date (MJD) ending by 4 or 9, and thus the period $T_{\mathrm{link/Yb}}$ in Eq.~(\ref{phasefreqdiffeq}) is consecutive days that are multiples of 5. Since $T_{\mathrm{Yb}}$ does not match $T_{\mathrm{link/Yb}}$ due to the dead time of NMIJ-Yb1, we extrapolate $y(\mathrm{Yb - UTC(NMIJ)})_{T_{\mathrm{Yb}}}$ to $y(\mathrm{Yb - UTC(NMIJ)})_{T_{\mathrm{link/Yb}}}$ by employing $y(\mathrm{Yb - UTC(NMIJ)})_{T_{\mathrm{Yb}}}$ as $y(\mathrm{Yb - UTC(NMIJ)})_{T_{\mathrm{link/Yb}}}$ and adding a correction frequency expected from a deterministic behavior of UTC(NMIJ) (see Section \ref{uncertaintyevaluationsection}). In 2023, the frequency of NMIJ-Yb1 was measured against a hydrogen maser (HM) and linked to UTC(NMIJ) by 
\begin{eqnarray}
y(\mathrm{Yb - UTC(NMIJ)})_{T_{\mathrm{link/Yb}}}&=&y(\mathrm{Yb - HM})_{T_{\mathrm{link/Yb}}}\nonumber\\
&&+ y(\mathrm{HM - UTC(NMIJ)})_{T_{\mathrm{link/Yb}}},
\end{eqnarray}
with $y(\mathrm{HM - UTC(NMIJ)})_{T_{\mathrm{link/Yb}}}$ calculated by BIPM from our measured time differences of 1 pulse-per-second signals between HM and UTC(NMIJ).

The calibration of TAI is carried out by calculating the mean frequency of TAI over a month referenced to an ensemble of primary and secondary frequency standards (PSFSs) via UTC($k$)s at the laboratory $k$. For this, BIPM accumulates the frequency evaluation results $y(\mathrm{TAI-PFS})_{T_{\mathrm{link/PFS}}}$ and $y(\mathrm{TAI-SFS})_{T_{\mathrm{link/SFS}}}$ by PSFSs including our result $y(\mathrm{TAI-Yb})_{T_{\mathrm{link/Yb}}}$, which are also given in Circular T. From these results, the frequency of NMIJ-Yb1 referenced to an individual PFS is derived by
\begin{equation}
y(\mathrm{Yb-PFS})_{T_{\mathrm{link}}} = y(\mathrm{Yb-TAI})_{T_{\mathrm{link}}}+y(\mathrm{TAI-PFS})_{T_{\mathrm{link}}}.
\label{ybpfseq}
\end{equation}
If $T_{\mathrm{link/Yb}}$ coincides with $T_{\mathrm{link/PFS}}$, the averaging period $T_{\mathrm{link}}$ in Eq.~(\ref{ybpfseq}) is the same as $T_{\mathrm{link/Yb}}$. If not, we extrapolate either $y(\mathrm{Yb-TAI})_{T_{\mathrm{link/Yb}}}$ or $y(\mathrm{PFS-TAI})_{T_{\mathrm{link/PFS}}}$ so that $T_{\mathrm{link/Yb}}$ matches $T_{\mathrm{link/PFS}}$, and set $T_{\mathrm{link}}$ as the extrapolated period. The maximum period of $T_{\mathrm{link}}$ is 25, 30, or 35 days for each month due to convention of the TAI calibration. The absolute frequency of Yb is derived from a weighted mean of individual values of $y(\mathrm{Yb-PFS})_{T_{\mathrm{link}}}$ with weights determined by an analysis described in Section \ref{analysissection}.

\subsection{Operation periods of the clocks}
\begin{figure}[h]
\includegraphics[scale=0.5]{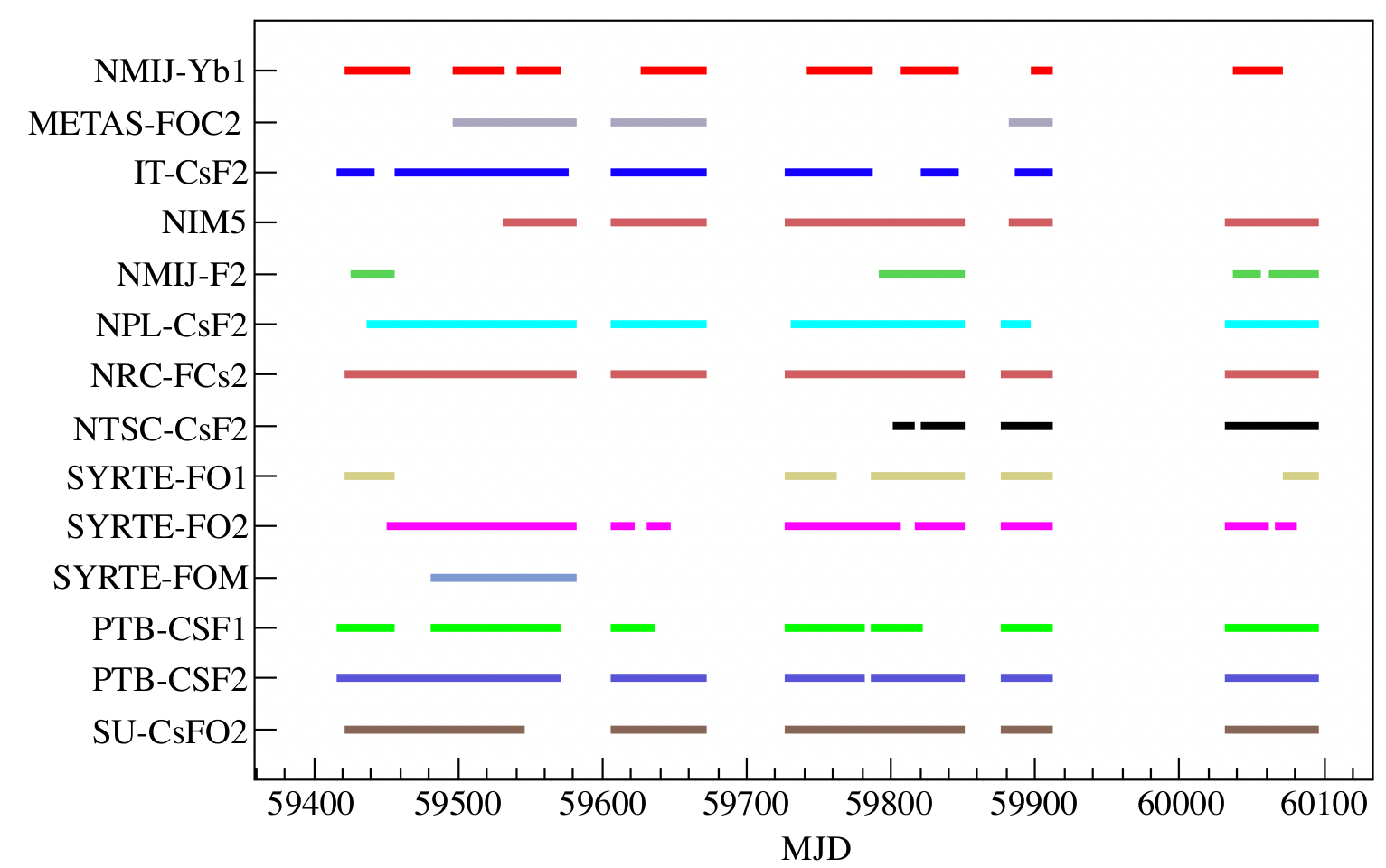}
\caption{Operation periods of NMIJ-Yb1 and those of 13 PFSs that are chosen to determine the absolute frequency of Yb.}
\label{operationrecord}
\end{figure}

NMIJ-Yb1 was operated from MJD 59424 (29 July 2021) to MJD 60069 (5 May 2023) and contributed to TAI evaluations for 14 months. Table \ref{yboperation} summarizes the operation periods of NMIJ-Yb1. The unique feature of NMIJ-Yb1 is the capability of the operation with a high uptime assisted by automation and remote systems \cite{Kobayashi2020,Kobayashi2019}, which contributes to reducing the uncertainty due to the extrapolation (see Section \ref{uncertaintyevaluationsection}). During the 14-month periods, TAI evaluation reports were submitted to BIPM from 13 Cs-fountain PFSs: METAS-FOC2 (METAS: Swiss Federal Institute of Metrology) \cite{Jallageas2018}, IT-CsF2 (IT: INRIM) \cite{Levi2014}, NIM5 (NIM: National Institute of Metrology of China) \cite{Fang2015}, NMIJ-F2, NPL-CsF2 (NPL: National Physical Laboratory) \cite{Li2011}, NRC-FCs2 (NRC: National Research Council Canada) \cite{Beattie2020}, NTSC-CsF2 (NTSC: National Time Service Center) \cite{Wang2023}, SYRTE-FO1, SYRTE-FO2, SYRTE-FOM (SYRTE: Syst\`{e}mes de R\'ef\'erence Temps-Espace) \cite{Guena2012}, PTB-CSF1, PTB-CSF2 (PTB: Physikalisch-Technische Bundesanstalt) \cite{Weyers2018}, and SU-CsFO2 (SU: All-Russian Scientific Research Institute of Physical Technical Measurements (VNIIFTRI)) \cite{Domnin2013}.  Figure \ref{operationrecord} shows the operation periods of these PFSs that are chosen to determine the absolute frequency of Yb. In total, we obtained 130 values of $y(\mathrm{Yb-PFS})_{T_{\mathrm{link}}}$. 

\begin{table}[t]
\caption{Operation records of NMIJ-Yb1 for TAI evaluations.}  
	\label{yboperation}
	\begin{center} 
\begin{tabular}{cccl}
\hline
MJD  & Periods (days) & Uptime ($\%$) & Issue month of Circular T\\
\hline
$59424-59454$  & 30 & 94.5 & August 2021 \\
$59454-59464$ & 10 & 94.3 & September 2021\\
$59499-59514$ & 15 & 95.3 & October 2021 \\
$59514-59529$ & 15 & 91.8 & November 2021\\
$59544-59569$ & 25 & 90.6 & December 2021\\
$59629-59634$ & 5 & 89.2 & February 2022\\
$59634-59654$ & 20 & 97.0 & March 2022$^{*}$\\
$59659-59669$ & 10 & 92.4 & March 2022$^{*}$\\
$59744-59759$ & 15 & 87.1 & June 2022\\
$59764-59784$ & 20 & 84.8 & July 2022\\
$59809-59819$ & 10 & 89.6 & August 2022\\
$59819-59844$ & 25 & 86.8 & September 2022\\
$59899-59909$ & 10 & 96.2 & November 2022\\
$60039-60064$ & 25 & 87.1 & April 2023\\
$60064-60069$ & 5 & 86.8 & May 2023\\
\hline
\end{tabular}
\end{center}
\small{$^{*}$ In March 2022, we separated the evaluation period into two parts, but combine the results of this month in the present analysis.}
\end{table}

\subsection{Uncertainty contributions}
\label{uncertaintyevaluationsection}
Similarly to the TAI evaluation \cite{circulart}, we take account of 11 uncertainty contributions for each measurement of $y(\mathrm{Yb-PFS})_{T_{\mathrm{link}}}$ arising from NMIJ-Yb1, PFS, and the link between them as follows. Typical values of these uncertainties are summarized in Table \ref{typicalupsfs}.

\begin{table}[t]
\caption{Examples of the uncertainties evaluated in each month, which are basically taken from Circular T \cite{circulart}. $u_{\mathrm{EAL}}^{\mathrm{Yb,Clock}}$ is calculated for $T_{\mathrm{link/Yb}}=30$ days from MJD 59424 to MJD 59454 (August 2021).}  
	\label{typicalupsfs}
	\begin{center} 
\begin{tabular}{lclllllll}
\hline
Clock & MJD & $u^{\mathrm{Clock}}_{\mathrm{A}}$ & $u^{\mathrm{Clock}}_{\mathrm{B}}$ & $u_{\mathrm{A/Lab}}^{\mathrm{Clock}}$ & $u_{\mathrm{B/Lab}}^{\mathrm{Clock}}$ & $u_{\mathrm{l/Tai}}^{\mathrm{Clock}}$ & $u_{\mathrm{EAL}}^{\mathrm{Yb,Clock}}$\\ 
&&\scriptsize{($\times10^{-15}$)}&\scriptsize{($\times10^{-15}$)}&\scriptsize{($\times10^{-15}$)}&\scriptsize{($\times10^{-15}$)}& \scriptsize{($\times10^{-15}$)} & \scriptsize{($\times10^{-15}$)}  \\
\hline
NMIJ-Yb1 & $59424-59454$ & 0.01 & $0.11^{*}$ & 0.04 & 0.10&0.20 & $-$\\
METAS-FOC2 & $59499-59514$ & 0.15 & 1.37 & 0.16 & 0.08  & 0.37 & $-$\\
IT-CsF2 & $59419-59439$ &0.21 &0.35 &0.10 &0.01& 0.28 &0.30 \\  
NIM5 & $59534-59544$ & 0.67 & 0.90 & 0.10 & 0.01  & 1.23 &$-$\\
NMIJ-F2 & $59429-59454$ &0.17 &  0.46  & 0.28 & 0.00  & 0.23 &0.13\\
NPL-CsF2 & $59439-59454$ & 0.21 &  0.20  & 0.13 & 0.05  & 0.37 &  0.30\\
NRC-FCs2 & $59424-59454$ & 0.11 & 0.21  & 0.10 & 0.00  & 0.20 & 0.00\\
NTSC-CsF2 & $59804-59814$ & 0.69 & 0.51 & 0.10 & 0.00  & 0.53 &$-$\\
SYRTE-FO1 & $59424-59454$ & 0.20 & 0.33  & 0.10 & 0.00 & 0.20 & 0.00\\
SYRTE-FO2 & $59454-59484$ & 0.20 &  0.22 & 0.09 & 0.00 & 0.20 & $-$\\
SYRTE-FOM & $59484-59514$  & 0.20 &  0.56 & 0.05 & 0.00  & 0.20 & $-$\\
PTB-CSF1 & $59419-59454$ & 0.07 &  0.34  & 0.01 & 0.00  & 0.06  & 0.11\\
PTB-CSF2 & $59419-59454$ & 0.09 &  0.17  & 0.07 & 0.00  & 0.06  & 0.11\\
SU-CsFO2 & $59424-59454$ & 0.25 & 0.22  & 0.11 & 0.00  & 0.20 & 0.00\\	  
\hline
\end{tabular}
\end{center}
\small{$^{*}$$u^{\mathrm{Yb}}_{\mathrm{B}}$ is slightly different from the value in Circular T, since we have updated some systematic evaluations \cite{Kobayashi2022,Nakashima2022} after submitting the TAI evaluation report to BIPM.}

\end{table}

\begin{table}[h]
\caption{Example of the systematic uncertainty budget of NMIJ-Yb1 \cite{Kobayashi2022}. This budget is evaluated for measurements in October 2021. BBR: blackbody radiation, AOM: acousto-optic modulator.}  
	\label{systematictalbe}
	\begin{center} 
\begin{tabular}{lcc}
\hline
Effect  & $\mathrm{Shift}$ $(\times10^{-17}$) & $\mathrm{Uncertainty}$ $(\times10^{-17}$) \\
\hline
Lattice light & 6.2 & 5.1 \\
BBR & $-252.3$ & 8.0 \\
Density & $-2.4$ & 1.2 \\
Second-order Zeeman & $-5.0$ & 0.3 \\
Probe light & 0.4 & 1.0 \\
Servo error & $-2.4$ & 1.1\\
AOM switching & $-$& 1 \\
Line pulling & $-$& 1 \\
DC Stark & $-$& $0.1$ \\
Gravitational redshift &230.8 &0.6\\
\hline
Total $u^{\mathrm{Yb}}_{\mathrm{B}}$ & $-24.8$ & 9.8$^{*}$\\
\hline
\end{tabular}
\end{center}
\small{$^{*}$$u^{\mathrm{Yb}}_{\mathrm{B}}$ is slightly different from the value in Circular T, since we have updated some systematic evaluations \cite{Kobayashi2022,Nakashima2022} after submitting the TAI evaluation report to BIPM.}
\end{table}

The statistical uncertainty $u^{\mathrm{Yb}}_{\mathrm{A}}$ of NMIJ-Yb1 is estimated as $\lesssim1\times10^{-17}$ by $u^{\mathrm{Yb}}_{\mathrm{A}}(\tau)=1.0\times10^{-14}/\sqrt{(\tau/\mathrm{s})}$ where $\tau$ denotes an averaging time, which is evaluated by our local Yb/Sr frequency ratio measurement \cite{Hisai2021}.

The systematic uncertainty $u^{\mathrm{Yb}}_{\mathrm{B}}$ of NMIJ-Yb1 is estimated as $\sim1\times10^{-16}$. Table \ref{systematictalbe} shows a typical uncertainty budget of NMIJ-Yb1 in the present measurement, based on the systematic evaluation described in Ref.~\cite{Kobayashi2022}. Note that the uncertainty of the probe light shift is increased compared with that of Ref.~\cite{Kobayashi2022}, taking account of recent studies on the effect of small residual ellipticity of the probe laser \cite{Yudin2023}. The gravitational redshift is evaluated with respect to the equipotential $W_{0}=$62 636 856.0 m$^{2}$/s$^{2}$ which is conventionally adopted for the transformation of the proper time of a frequency standard to TAI \cite{CGPM2018}. Compared with our previous measurement \cite{Kobayashi2020}, the uncertainty of the gravitational redshift is improved thanks to reevaluation of the geopotential of NMIJ-Yb1 by Geospatial Information Authority of Japan \cite{Nakashima2022}. 

The statistical uncertainty $u_{\mathrm{A/Lab}}^{\mathrm{Yb}}$ of the link between NMIJ-Yb1 and UTC(NMIJ) is estimated as $\lesssim2\times10^{-16}$, arising from the extrapolation of $y(\mathrm{Yb - LO})_{T_{\mathrm{Yb}}}$ to $y(\mathrm{Yb - LO})_{T_{\mathrm{link/Yb}}}$, where LO is UTC(NMIJ) or HM. This uncertainty is estimated by a numerical simulation of the stochastic behavior of a local oscillator \cite{Yu2007}. When NMIJ-Yb1 is directly linked to UTC(NMIJ) from 2021 to 2022, we simulate time series data of the frequency according to a noise model of UTC(NMIJ) that approximately characterizes the Allan deviation of UTC(NMIJ) \cite{Kobayashi2020}: $1\times10^{-12}/(\tau/\mathrm{s})$ for the white phase modulation (PM), $9\times10^{-14}/\sqrt{(\tau/\mathrm{s})}$ for the white frequency modulation (FM), $2\times10^{-15}$ for the flicker FM, and $4\times10^{-24}\sqrt{(\tau/\mathrm{s})}$ for the random walk FM. We generate 200 time series data and calculate the root-mean-square deviation $\sqrt{\overline{(y^{\mathrm{LO}}_{T_{\mathrm{link/Yb}}}-y^{\mathrm{LO}}_{T_{\mathrm{Yb}}})^{2}}}$ between simulated frequencies $y^{\mathrm{LO}}_{T_{\mathrm{link/Yb}}}$ and $y^{\mathrm{LO}}_{T_{\mathrm{Yb}}}$ averaged over $T_{\mathrm{link/Yb}}$ and $T_{\mathrm{Yb}}$, respectively. The obtained $\sqrt{\overline{(y^{\mathrm{LO}}_{T_{\mathrm{link/Yb}}}-y^{\mathrm{LO}}_{T_{\mathrm{Yb}}})^{2}}}$ is assigned as the uncertainty arising from the extrapolation, which is added to $u_{\mathrm{A/Lab}}^{\mathrm{Yb}}$. When the frequency steering of UTC(NMIJ) is performed during $T_{\mathrm{link/Yb}}$, the dead time of NMIJ-Yb1 causes a deterministic frequency shift of $y(\mathrm{Yb - UTC(NMIJ)})_{T_{\mathrm{Yb}}}$ from $y(\mathrm{Yb - UTC(NMIJ)})_{T_{\mathrm{link/Yb}}}$. We correct this shift and include the uncertainty of the shift in $u_{\mathrm{A/Lab}}^{\mathrm{Yb}}$. When NMIJ-Yb1 is compared with HM in 2023, we use a noise model of HM \cite{Kobayashi2024}: $3\times10^{-13}/(\tau/\mathrm{s})$ for the white PM, $6\times10^{-14}/\sqrt{(\tau/\mathrm{s})}$ for the white FM, $5\times10^{-16}$ for the flicker FM, and $2\times10^{-27}\sqrt{(\tau/\mathrm{s})}$ for the random walk FM. Since HM exhibits a linear frequency drift of $-1\times10^{-16}$/d referenced to NMIJ-Yb1, we correct a deterministic frequency shift of $y(\mathrm{Yb - HM})_{T_{\mathrm{Yb}}}$ from $y(\mathrm{Yb - HM})_{T_{\mathrm{link/Yb}}}$ due to the dead time of NMIJ-Yb1, and include the uncertainty of this shift in $u_{\mathrm{A/Lab}}^{\mathrm{Yb}}$. Note that the effect of the linear drift of UTC(NMIJ) is not considered, since the linear drift referenced to TAI \cite{circulart} is negligibly small.

The systematic uncertainty $u_{\mathrm{B/Lab}}^{\mathrm{Yb}}$ of the link between NMIJ-Yb1 and UTC(NMIJ) is estimated as $1.0\times10^{-16}$, dominated by the effect of phase variations of the 10 MHz signal during its transmission through a coaxial cable between NMIJ-Yb1 and UTC(NMIJ) \cite{Kobayashi2022}. In our previous measurement \cite{Kobayashi2020}, $u_{\mathrm{B/Lab}}^{\mathrm{Yb}}$ was $2.2\times10^{-16}$ limited by phase variations of the RF signal in a frequency multiplier. The contribution from this effect is negligible here by stabilizing the temperature of the frequency multiplier \cite{Wada2022}.

The uncertainty $u_{\mathrm{l/Tai}}^{\mathrm{Yb}}$ of the GNSS link to TAI is typically several $10^{-16}$, which is calculated by the recommended formula \cite{Panfilo2010}
\begin{equation}
u_{\mathrm{l/Tai}}^{\mathrm{Yb}}(T_{\mathrm{link/Yb}})=\frac{\sqrt{u_{\mathrm{A\,\mathrm{i}}}^{2}+u_{\mathrm{A\,\mathrm{f}}}^{2}}}{432000[\frac{(T_{\mathrm{link/Yb}}/\mathrm{d})}{5}]^{0.9}},
\label{linkeq}
\end{equation}
with the statistical uncertainties $u_{\mathrm{A\,\mathrm{i}}}$ and $u_{\mathrm{A\,\mathrm{f}}}$ ($=0.3$ or 0.4 ns) of the time difference measurements of $\Delta t_{\mathrm{i}}$ and $\Delta t_{\mathrm{f}}$ in Eq.~(\ref{phasefreqdiffeq}), respectively. 

The corresponding uncertainties $u_{\mathrm{A}}^{\mathrm{PFS}}$, $u_{\mathrm{B}}^{\mathrm{PFS}}$, $u_{\mathrm{A/Lab}}^{\mathrm{PFS}}$, $u_{\mathrm{B/Lab}}^{\mathrm{PFS}}$, $u_{\mathrm{l/Tai}}^{\mathrm{PFS}}$ for PFS are taken from Circular T. 

The uncertainty $u_{\mathrm{EAL}}^{\mathrm{Yb,PFS}}$ due to the extrapolation of $y(\mathrm{Clock-TAI})_{T_{\mathrm{link/Clock}}}$ to $y(\mathrm{Clock-TAI})_{T_{\mathrm{link}}}$, where Clock is Yb or PFS, is estimated as typically several $10^{-16}$, which is calculated by a numerical simulation based on a noise model of Echelle Atomique Libre (EAL) given in Circular T: $1\times10^{-15}/\sqrt{(\tau/\mathrm{d})}$ for the white FM, $2\times10^{-16}$ for the flicker FM, and $2\times10^{-17}\sqrt{(\tau/\mathrm{d})}$ for the random walk FM. Similarly to the estimation of $u_{\mathrm{A/Lab}}^{\mathrm{Yb}}$, we derive the root-mean square deviation $\sqrt{\overline{(y^{\mathrm{EAL}}_{T_{\mathrm{link/Clock}}}-y^{\mathrm{EAL}}_{T_{\mathrm{link}}})^{2}}}$ between simulated frequencies $y^{\mathrm{EAL}}_{T_{\mathrm{link/Clock}}}$ and $y^{\mathrm{EAL}}_{T_{\mathrm{link}}}$ averaged over $T_{\mathrm{link/Clock}}$ and $T_{\mathrm{link}}$, respectively, and assign this as $u_{\mathrm{EAL}}^{\mathrm{Yb,PFS}}$.

\subsection{Analysis to derive the weighted mean}
\label{analysissection}
Following an analysis method with covariance matrices described in Ref.~\cite{Nemitz2021}, we derive the weighted mean of the 130 individual measurement values of $y(\mathrm{Yb-PFS})_{T_{\mathrm{link}}}$. The weighted mean $\overline{y(\mathrm{Yb-PFS})}$ is described by 
\begin{equation}
\overline{y(\mathrm{Yb-PFS})}=\boldsymbol{w}^{\mathrm{T}}\boldsymbol{y},
\label{weightedmeaneq}
\end{equation}
where $\boldsymbol{y}$ and $\boldsymbol{w}$ denote column vectors with the 130 values of $y(\mathrm{Yb-PFS})_{T_{\mathrm{link}}}$ and their normalized weights, respectively. The uncertainty $u_{\mathrm{total}}$ of $\overline{y(\mathrm{Yb-PFS})}$ is given by 
\begin{equation}
u_{\mathrm{total}}^{2}=\boldsymbol{w}^{\mathrm{T}}C_{u_{\mathrm{total}}}\boldsymbol{w},
\label{utotaleq}
\end{equation}
where $C_{u_{\mathrm{total}}}$ is a $130\times130$ covariance matrix, which is calculated by a sum of individual covariance matrices, denoted by $C_{u}$ for the uncertainty type $u$, of the 11 uncertainty contributions described in Section \ref{uncertaintyevaluationsection}, and 2 negative contributions due to shared link uncertainties of NMIJ-Yb1 and NMIJ-F2, 
\begin{eqnarray}
C_{u_{\mathrm{total}}}&=&\sum_{u}C_{u}\nonumber\\
&=&C_{u_{\mathrm{A}}^{\mathrm{Yb}}}+C_{u_{\mathrm{B}}^{\mathrm{Yb}}}+C_{u_{\mathrm{A/Lab}}^{\mathrm{Yb}}}+C_{u_{\mathrm{B/Lab}}^{\mathrm{Yb}}}+C_{u_{\mathrm{l/Tai}}^{\mathrm{Yb}}}\nonumber\\
&&+C_{u_{\mathrm{A}}^{\mathrm{PFS}}}+C_{u_{\mathrm{B}}^{\mathrm{PFS}}}+C_{u_{\mathrm{A/Lab}}^{\mathrm{PFS}}}+C_{u_{\mathrm{B/Lab}}^{\mathrm{PFS}}}+C_{u_{\mathrm{l/Tai}}^{\mathrm{PFS}}}\nonumber\\
&&+C_{u_{\mathrm{EAL}}^{\mathrm{Yb,PFS}}}-C_{u_{\mathrm{A/Lab}}^{\mathrm{Yb,NMIJ\mathchar`-F2}}}-C_{u_{\mathrm{l/Tai}}^{\mathrm{Yb,NMIJ\mathchar`-F2}}}.
\label{ctotaleq}
\end{eqnarray}
$C_{u}$ except for $C_{u_{\mathrm{A/Lab}}^{\mathrm{Yb,NMIJ\mathchar`-F2}}}$ and $C_{u_{\mathrm{l/Tai}}^{\mathrm{Yb,NMIJ\mathchar`-F2}}}$ is derived from a $130\times130$ diagonal matrix $D_{u}$ including 130 uncertainties of the type $u$ and a $130\times130$ correlation matrix $R_{u}$,
\begin{equation}
C_{u}=D_{u}R_{u}D_{u}.
\end{equation}
$C_{u_{\mathrm{A/Lab}}^{\mathrm{Yb,NMIJ\mathchar`-F2}}}$ ($C_{u_{\mathrm{l/Tai}}^{\mathrm{Yb,NMIJ\mathchar`-F2}}}$) is a sparse matrix with diagonal elements including an amount to be subtracted from a sum of variance $(u_{\mathrm{A/Lab}}^{\mathrm{Yb}})^{2}+(u_{\mathrm{A/Lab}}^{\mathrm{NMIJ\mathchar`-F2}})^{2}$ ($(u_{\mathrm{l/Tai}}^{\mathrm{Yb}})^{2}+(u_{\mathrm{l/Tai}}^{\mathrm{NMIJ\mathchar`-F2}})^{2}$), taking account of negative correlations in the laboratory (GNSS) link uncertainties. Once $C_{u_{\mathrm{total}}}$ in Eq.~(\ref{ctotaleq}) is obtained, we find $\boldsymbol{w}$ which minimizes $u_{\mathrm{total}}^{2}$ in Eq.~(\ref{utotaleq}) with constraints (a) $\boldsymbol{j}^{\mathrm{T}}\boldsymbol{w}=1$ where $\boldsymbol{j}$ is an all-ones column vector and (b) that all elements of $\boldsymbol{w}$ are non-negative. This optimization problem is solved by a code written in Mathematica.

$C_{u_{\mathrm{A/Lab}}^{\mathrm{Yb,NMIJ\mathchar`-F2}}}$ arises from an overlapping period in the extrapolated periods $T_{\mathrm{link/Yb}}$ and $T_{\mathrm{link/NMIJ\mathchar`-F2}}$ by the same hydrogen maser. Non-zero elements of $C_{u_{\mathrm{A/Lab}}^{\mathrm{Yb,NMIJ\mathchar`-F2}}}$ are approximately estimated by $2u_{\mathrm{A/Lab}}^{\mathrm{Yb}}u_{\mathrm{A/Lab}}^{\mathrm{NMIJ\mathchar`-F2}}\sqrt{T_{\mathrm{over}}^{2}/(T_{\mathrm{link/Yb}}T_{\mathrm{link/NMIJ\mathchar`-F2}})}$, where $T_{\mathrm{over}}$ denotes the overlapping period between $T_{\mathrm{link/Yb}}$ and $T_{\mathrm{link/NMIJ\mathchar`-F2}}$. Note that a more exact estimation of $C_{u_{\mathrm{A/Lab}}^{\mathrm{Yb,NMIJ\mathchar`-F2}}}$ requires the uptimes of NMIJ-Yb1 and NMIJ-F2 and the noise model of the hydrogen maser, but we here employ approximate values due to the high uptime operations of these clocks. $C_{u_{\mathrm{l/Tai}}^{\mathrm{Yb,NMIJ\mathchar`-F2}}}$ takes account of the fact that the initial or final date of $T_{\mathrm{link/Yb}}$ coincides with that of $T_{\mathrm{link/NMIJ\mathchar`-F2}}$, and thus $y(\mathrm{Yb-TAI})_{T_{\mathrm{link/Yb}}}$ and $y(\mathrm{NMIJ\mathchar`-F2-TAI})_{T_{\mathrm{link/NMIJ\mathchar`-F2}}}$ share the common GNSS uncertainty $u_{\mathrm{A\,com}}=u_{\mathrm{A\,\mathrm{i}}}$ or $u_{\mathrm{A\,\mathrm{f}}}$. From Eq.~(\ref{linkeq}), non-zero elements of $C_{u_{\mathrm{l/Tai}}^{\mathrm{Yb,NMIJ\mathchar`-F2}}}$ are estimated by $2u_{\mathrm{A\,com}}^{2}/\{432000^{2}[\frac{(T_{\mathrm{link/NMIJ\mathchar`-F2}}/\mathrm{d})}{5}]^{0.9}[\frac{(T_{\mathrm{link/Yb}}/\mathrm{d})}{5}]^{0.9}\}$.

To determine matrix elements of $R_{u}$, we first consider correlations of the statistical uncertainties $u_{\mathrm{A}}^{\mathrm{Clock}}$, $u_{\mathrm{A/Lab}}^{\mathrm{Clock}}$, $u_{\mathrm{l/Tai}}^{\mathrm{Clock}}$, and $u_{\mathrm{EAL}}^{\mathrm{Yb,PFS}}$ between the 130 individual measurements of $y(\mathrm{Yb-PFS})_{T_{\mathrm{link}}}$ as follows. We consider no correlations of $u_{\mathrm{A}}^{\mathrm{PFS}}$ over all measurements. $u_{\mathrm{A}}^{\mathrm{Yb}}$ and $u_{\mathrm{A/Lab}}^{\mathrm{Yb}}$ are considered to be uncorrelated across temporally separated measurements. $u_{\mathrm{A/Lab}}^{\mathrm{PFS}}$ is correlated for PFSs in the same institutes when there is an overlapping period $T_{\mathrm{over}}$ in the extrapolated periods $T_{\mathrm{link1/PFS1}}$ and $T_{\mathrm{link2/PFS2}}$ by same hydrogen masers. In this case, the correlation coefficient is approximately estimated by $\sqrt{T_{\mathrm{over}}^{2}/(T_{\mathrm{link1/PFS1}}T_{\mathrm{link2/PFS2}})}$. $u_{\mathrm{l/Tai}}^{\mathrm{Clock}}$ is negatively correlated when temporally adjacent measurements share the same $u_{\mathrm{A\,com}}=u_{\mathrm{A\,\mathrm{i}}}$ or $u_{\mathrm{A\,\mathrm{f}}}$. The correlation coefficient $r_{u_{\mathrm{l/Tai}}^{\mathrm{Clock}}}$ for adjacent measurements with periods $T_{\mathrm{link1/Clock}}$ and $T_{\mathrm{link2/Clock}}$ is given by 
\begin{eqnarray}
r_{u_{\mathrm{l/Tai}}^{\mathrm{Clock}}}&=&-\frac{1}{u_{\mathrm{l/Tai}}^{\mathrm{Clock}}(T_{\mathrm{link1/Clock}})u_{\mathrm{l/Tai}}^{\mathrm{Clock}}(T_{\mathrm{link2/Clock}})}\nonumber\\
&&\times\frac{u_{\mathrm{A\,com}}^{2}}{432000^{2}[\frac{(T_{\mathrm{link1/Clock}}/\mathrm{d})}{5}]^{0.9}[\frac{(T_{\mathrm{link2/Clock}}/\mathrm{d})}{5}]^{0.9}},
\label{correlationltaieq}
\end{eqnarray}
which is $-0.5$ in most cases due to the fact that $u_{\mathrm{A\,\mathrm{i}}}=u_{\mathrm{A\,\mathrm{f}}}$. In addition, $u_{\mathrm{l/Tai}}^{\mathrm{PFS}}$ is positively correlated when PFSs in the same institutes share the same $u_{\mathrm{A\,com}}$. The correlation coefficient for this case is estimated in a similar way to Eq.~(\ref{correlationltaieq}) as 1 (0.5) when these PFSs share both $u_{\mathrm{A\,\mathrm{i}}}$ and $u_{\mathrm{A\,\mathrm{f}}}$ (either $u_{\mathrm{A\,\mathrm{i}}}$ or $u_{\mathrm{A\,\mathrm{f}}}$). $u_{\mathrm{EAL}}^{\mathrm{Yb,PFS}}$ is correlated when there is an overlapping period in the extrapolated periods $T_{\mathrm{link1}}$ and $T_{\mathrm{link2}}$ performed in different measurements $y(\mathrm{TAI-Clock1})_{T_{\mathrm{link/Clock1}}}$ and $y(\mathrm{TAI-Clock2})_{T_{\mathrm{link/Clock2}}}$, respectively (Clock1 and Clock2 denote Yb or PFS). In this case, we assign a correlation coefficient $r_{u_{\mathrm{EAL}}^{\mathrm{Yb,PFS}}}$ estimated by 
\begin{equation}
r_{u_{\mathrm{EAL}}^{\mathrm{Yb,PFS}}}=\frac{\overline{(y^{\mathrm{EAL}}_{T_{\mathrm{link/Clock1}}}-y^{\mathrm{EAL}}_{T_{\mathrm{link1}}})(y^{\mathrm{EAL}}_{T_{\mathrm{link/Clock2}}}-y^{\mathrm{EAL}}_{T_{\mathrm{link2}}})}}{\sqrt{\overline{(y^{\mathrm{EAL}}_{T_{\mathrm{link/Clock1}}}-y^{\mathrm{EAL}}_{T_{\mathrm{link1}}})^{2}}}\sqrt{\overline{(y^{\mathrm{EAL}}_{T_{\mathrm{link/Clock2}}}-y^{\mathrm{EAL}}_{T_{\mathrm{link2}}})^{2}}}},
\end{equation}
using a numerical simulation based on the noise model of EAL described in Section \ref{uncertaintyevaluationsection}. 

Regarding correlations of the systematic uncertainties $u_{\mathrm{B}}^{\mathrm{Clock}}$ and $u_{\mathrm{B/Lab}}^{\mathrm{Clock}}$, an assumption of strong correlations (i.e., correlation coefficients of 1) among the evaluations for the same clocks induces concentration of weights on measurements with lower systematic uncertainties, causing an uneven weight distribution for $y(\mathrm{Yb-PFS})_{T_{\mathrm{link}}}$. To avoid this, a previous work \cite{Nemitz2021} temporarily assigns a reduced correlation coefficient of 0.5, taking account of the possibility that the systematic uncertainty also includes statistical components that vary with time and so are averaged out. Here we similarly assign correlation coefficients of 0.5 for the same clocks to derive $\boldsymbol{w}$. When calculating the uncertainty with the derived $\boldsymbol{w}$, we conservatively assign correlation coefficients of 1. Regarding the correlations between different PFSs, correlation coefficients of 0 are assigned assuming no correlations in the systematic evaluations among different PFSs \cite{Nemitz2021,McGrew2019,Schwarz2020,Lange2021}.

\section{Results}
\begin{table}[h]
\caption{Uncertainty budget for the absolute frequency measurement of $^{171}$Yb.}  
	\label{absbduget}
	\begin{center} 
\begin{tabular}{lc}
\hline
Effect  & Uncertainty ($\times10^{-16}$)\\
\hline
NMIJ-Yb1 & \\ 
Clock (statistics) $u_{\mathrm{A}}^{\mathrm{Yb}}$ &0.03 \\
Clock (systematics) $u_{\mathrm{B}}^{\mathrm{Yb}}$ & 1.06 \\
Local link to UTC(NMIJ) (statistics) $u_{\mathrm{A/Lab}}^{\mathrm{Yb}}$& 0.33\\
Local link to UTC(NMIJ) (systematics) $u_{\mathrm{B/Lab}}^{\mathrm{Yb}}$& 1.00\\
Satellite link to TAI $u_{\mathrm{l/Tai}}^{\mathrm{Yb}}$ & 0.70\\
\hline
Ensemble of PFSs & \\ 
Clock (statistics) $u_{\mathrm{A}}^{\mathrm{PFS}}$ & 0.23\\
Clock (systematics) $u_{\mathrm{B}}^{\mathrm{PFS}}$ &  0.80\\
Local link to UTC($k$) (statistics) $u_{\mathrm{A/Lab}}^{\mathrm{PFS}}$& 0.12\\
Local link to UTC($k$) (systematics) $u_{\mathrm{B/Lab}}^{\mathrm{PFS}}$& 0.05\\
Satellite link to TAI $u_{\mathrm{l/Tai}}^{\mathrm{PFS}}$ & 0.21\\
\hline
EAL extrapolation $u_{\mathrm{EAL}}^{\mathrm{Yb,PFS}}$ & 0.47\\
\hline
Total $u_{\mathrm{total}}$ & 1.93 \\
\hline
\end{tabular}
\end{center}
\end{table}

\begin{figure}[h]
\includegraphics[scale=0.45]{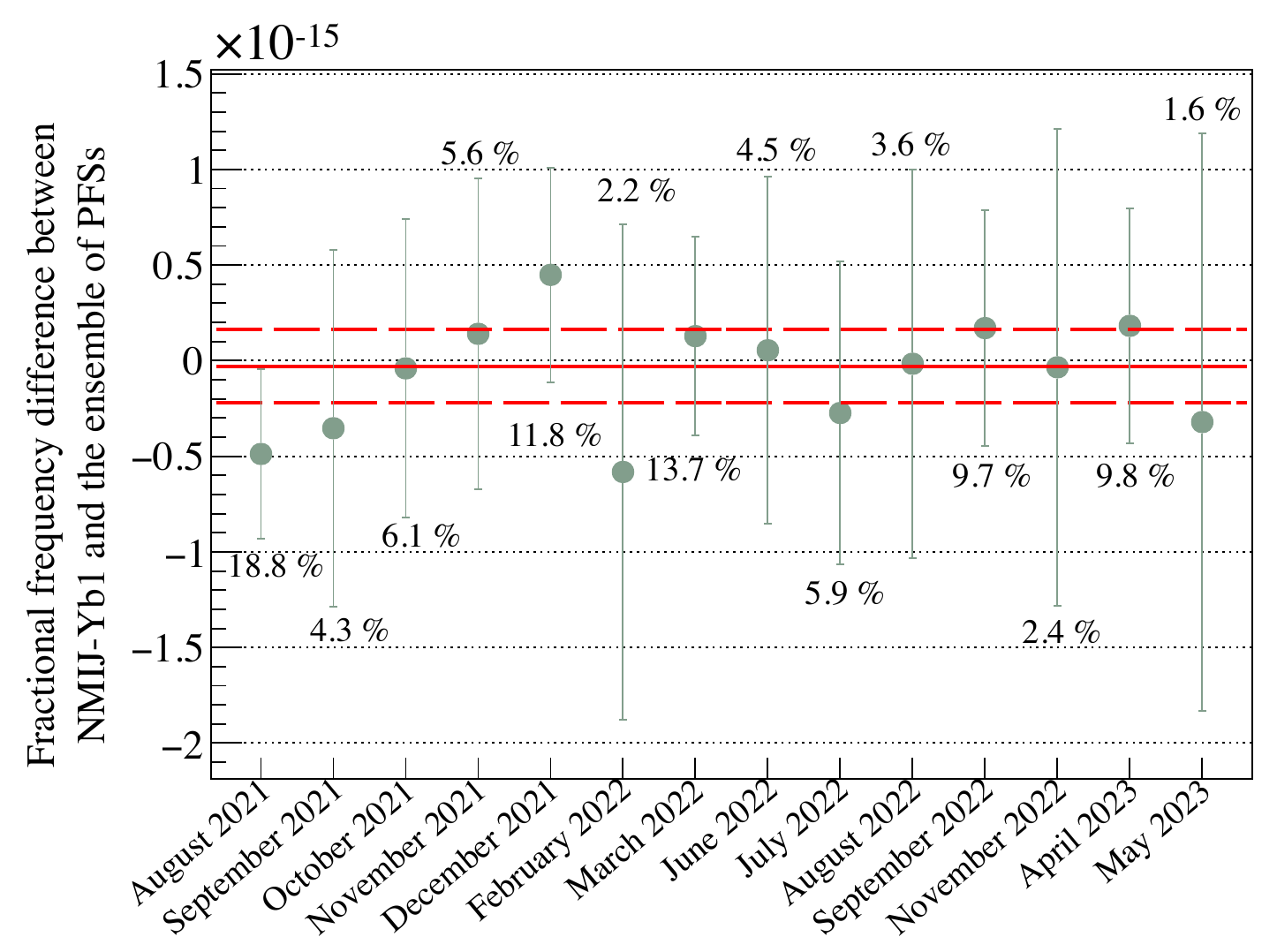}
\caption{Fractional frequency difference between NMIJ-Yb1 and the ensemble of PFSs calculated for each month. The error bar and the number expressed as a percentage indicate the total uncertainty and weight for each month, respectively. The solid and dashed red lines show the overall weighted mean $\overline{y(\mathrm{Yb-PFS})}=-0.29\times10^{-16}$ and its uncertainty $u_{\mathrm{total}}=1.93\times10^{-16}$ ($k=1$), respectively.}
\label{eachmonthfig}
\end{figure}

\begin{figure}[h]
\includegraphics[scale=0.45]{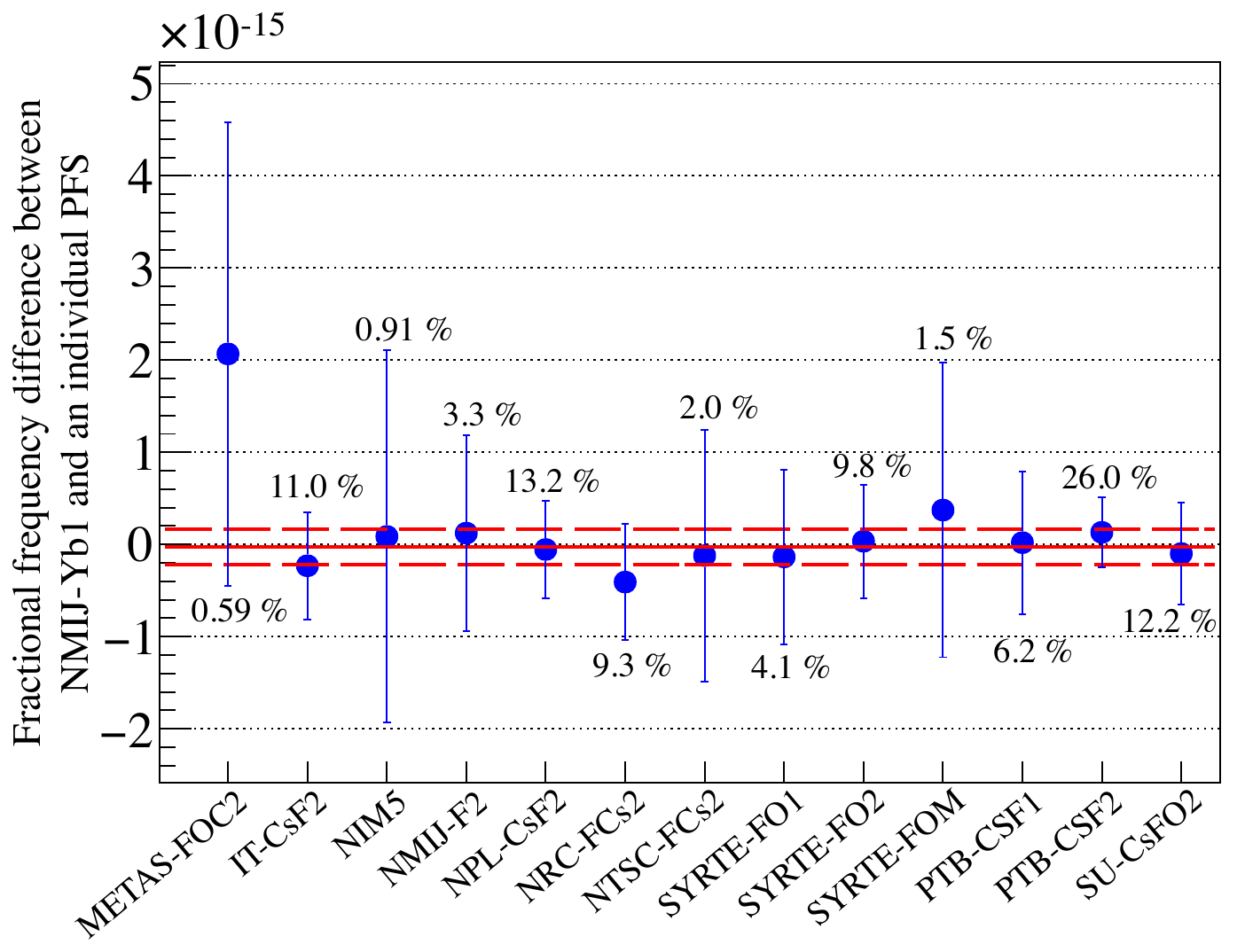}
\caption{Fractional frequency difference between NMIJ-Yb1 and an individual PFS averaged over the whole 14-month period. The error bar and the number expressed as a percentage indicate the total uncertainty and weight for each PFS, respectively. The solid and dashed red lines show the overall weighted mean $\overline{y(\mathrm{Yb-PFS})}=-0.29\times10^{-16}$ and its uncertainty $u_{\mathrm{total}}=1.93\times10^{-16}$ ($k=1$), respectively.}
\label{eachpfsfig}
\end{figure}

\noindent With obtained $\boldsymbol{w}$ and Eq.~(\ref{utotaleq}), we find $u_{\mathrm{total}}=1.93\times10^{-16}$. The uncertainty of the individual contribution is given by $\sqrt{\boldsymbol{w}^{\mathrm{T}}C_{u}\boldsymbol{w}}$, which is summarized in Table \ref{absbduget}. $u_{\mathrm{total}}$ is mostly limited by $u_{\mathrm{B}}^{\mathrm{Yb}}$ and $u_{\mathrm{B/Lab}}^{\mathrm{Yb}}$. Thanks to the contributions from 13 PFSs, $u_{\mathrm{B}}^{\mathrm{PFS}}$ is reduced to $0.80\times10^{-16}$, lower than $u_{\mathrm{B}}^{\mathrm{Yb}}$.

From Eq.~(\ref{weightedmeaneq}), we obtain $\overline{y(\mathrm{Yb-PFS})}=-0.29(1.93)\times10^{-16}$. The absolute frequency of Yb is determined as 518 295 836 590 863.62(10) Hz. We also calculate a sum of weights for each month or an individual PFS which is useful for calculating correlation coefficients between our present measurement and other measurements (see Section \ref{correlationssection}). This calculation yields the weighted mean values over contributions from each month and an individual PFS, which are shown in Figures \ref{eachmonthfig} and \ref{eachpfsfig}, respectively.

\begin{figure}[h]
\includegraphics[scale=0.46]{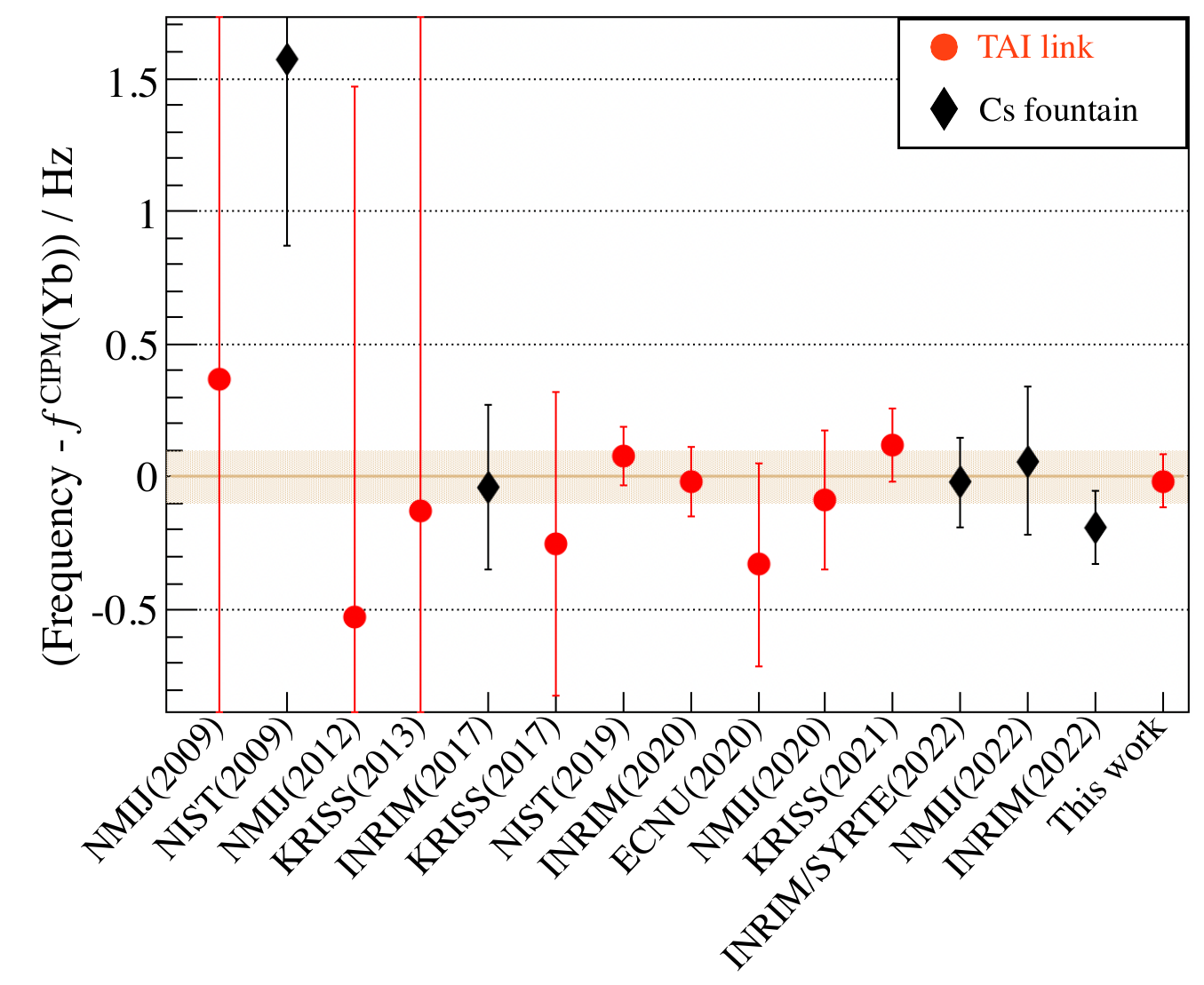}
\caption{Absolute frequencies of $^{171}$Yb measured using the TAI link \cite{Kohno2009,Yasuda2012,Park2013,Kim2017,McGrew2019,Pizzocaro2019,Luo2020,Kobayashi2020,Kim2021} and Cs fountain clocks \cite{Lemke2009,Pizzocaro2017,Kobayashi2022,Clivati2022,Goti2023}. The shaded region shows the CIPM recommended frequency $f^{\mathrm{CIPM}}(\mathrm{Yb})=$ 518 295 836 590 863.63 Hz with a fractional uncertainty of $1.9\times10^{-16}$ updated in 2021 \cite{Margolis2024}. ECNU: East China Normal University.}
\label{previousdata}
\end{figure}

\section{Discussions}
\label{discussionsection}
\subsection{Comparisons with previous absolute frequency measurements}
Our determined absolute frequency is in good agreement with the recommended frequency $f^{\mathrm{CIPM}}(\mathrm{Yb})$ and the results of previous absolute frequency measurements by several groups, as shown in Figure \ref{previousdata}. Our uncertainty $u_{\mathrm{total}}$ is comparable to the uncertainty of $f^{\mathrm{CIPM}}(\mathrm{Yb})$ ($1.9\times10^{-16}$) and slightly better than those of recent measurements by NIST in 2019 ($2.1\times10^{-16}$ \cite{McGrew2019}), INRIM in 2020 ($2.6\times10^{-16}$ \cite{Pizzocaro2019}), and KRISS in 2021 ($2.6\times10^{-16}$ \cite{Kim2021}). 

Regarding the other optical transitions, the uncertainty below $2\times10^{-16}$ has been reported for the $^{1}$S$_{0}-^{3}$P$_{0}$ transition of $^{87}$Sr ($1.5\times10^{-16}$ \cite{Schwarz2020} and $1.8\times10^{-16}$ \cite{Nemitz2021}) and the $^{2}$S$_{1/2}-^{2}$F$_{7/2}$ transition of $^{171}$Yb$^{+}$ ($1.3\times10^{-16}$ \cite{Lange2021}). While the lowest uncertainty of $1.3\times10^{-16}$ is achieved by local measurements with PTB-CSF1 and PTB-CSF2, a measurement using the TAI link also potentially reaches this level or below by taking advantage of the use of many PFSs. For example, the uncertainty of our measurement would be $1.3\times10^{-16}$ if we improve both $u_{\mathrm{B}}^{\mathrm{Yb}}$ and $u_{\mathrm{B/Lab}}^{\mathrm{Yb}}$ to $3\times10^{-17}$ (see Table \ref{absbduget}).

\subsection{Correlations between our present and previous measurements}
\label{correlationssection}
\begin{table}[t]
\caption{Absolute frequency measurements of $^{171}$Yb and a local frequency ratio measurement of $^{171}$Yb/$^{87}$Sr involving NMIJ-Yb1.}  
	\label{measurementtable}
	\begin{center} 
\begin{tabular}{ccccc}
\hline
Atomic species & Label  & Measured result & Fractional uncertainty &  Ref. \\
\hline
$^{171}$Yb & $m_{1}$ & 518 295 836 590 863.62(10) Hz & $u_{\mathrm{total}}^{m_{1}}=1.9\times10^{-16}$  & This work  \\
$^{171}$Yb & $m_{2}$ & 518 295 836 590 863.69(28) Hz & $u_{\mathrm{total}}^{m_{2}}=5.3\times10^{-16}$ & \cite{Kobayashi2022}  \\
$^{171}$Yb & $m_{3}$ & 518 295 836 590 863.54(26) Hz$^{*}$ & $u_{\mathrm{total}}^{m_{3}}=5.0\times10^{-16}$  & \cite{Kobayashi2020}  \\
$^{171}$Yb/$^{87}$Sr & $m_{4}$ & 1.207 507 039 343 338 58(49)$^{*}$ & $u_{\mathrm{total}}^{m_{4}}=4.1\times10^{-16}$  & \cite{Hisai2021}  \\
\hline
\end{tabular}
\end{center}
\small{$^{*}$The value included in the update of the CIPM 2021 recommended frequency \cite{Margolis2024}.}
\end{table}

\noindent At the next CCTF meeting, we plan to report the present result as a new input for updating the recommended frequency of $^{171}$Yb. Since the recommended frequency is calculated by combining measured results of absolute frequencies and frequency ratios \cite{Margolis2015}, it is important to take account of correlations among them \cite{Margolis2024}. Therefore, we estimate correlation coefficients between our present and previous measurements involving NMIJ-Yb1 that are labeled as $m_{i}$ with a total uncertainty $u_{\mathrm{total}}^{m_{i}}$ ($i=1,2,3,4$) in Table \ref{measurementtable}. Our estimated correlation coefficients $r_{m_{i},m_{j}}$ between measurements $m_{i}$ and $m_{j}$ ($i\neq j$) are summarized in Table \ref{corelationresults}. Their calculation methods are described below. 

First, we estimate the correlation coefficient $r_{m_{1},m_{2}}$ between our present measurement $m_{1}$ and the previous local measurement $m_{2}$ with NMIJ-F2. These two measurements share the common systematic uncertainties $u_{\mathrm{B}}^{\mathrm{Yb}}$, $u_{\mathrm{B}}^{\mathrm{\mathrm{NMIJ\mathchar`-F2}}}$, $u_{\mathrm{B/Lab}}^{\mathrm{Yb}}$ and statistical uncertainties $u_{\mathrm{A}}^{\mathrm{Yb}}$, $u_{\mathrm{A}}^{\mathrm{\mathrm{NMIJ\mathchar`-F2}}}$, $u_{\mathrm{A/Lab}}^{\mathrm{\mathrm{NMIJ\mathchar`-F2}}}$ due to a overlapping period $T_{m_{1},m_{2}}\sim18.47$ days in August 2021 among a total period $T_{\mathrm{m_{2}}}\sim45.89$ days of $m_{2}$. Note that the extrapolation procedure with the maser noise model was not performed in $m_{2}$, and so we do not consider a correlation arising from $u_{\mathrm{A/Lab}}^{\mathrm{Yb}}$ but only from an uncertainty $u_{\mathrm{cable}}^{\mathrm{NMIJ\mathchar`-F2}}=1.6\times10^{-16}$ included in $u_{\mathrm{A/Lab}}^{\mathrm{NMIJ\mathchar`-F2}}$, which was evaluated by measurements of the phase noise in cables \cite{circulart}. Following guidelines \cite{Margolisguideline}, the correlation coefficient is estimated by 
\begin{eqnarray} 
r_{m_{1},m_{2}} &&= \frac{1}{u_{\mathrm{total}}^{m_{1}}u_{\mathrm{total}}^{m_{2}}}[u_{\mathrm{B}}^{\mathrm{Yb}}u_{\mathrm{B}}^{\mathrm{Yb\,}m_{2}}+(u_{\mathrm{B/Lab}}^{\mathrm{Yb}})^{2}+w_{\mathrm{total}}^{\mathrm{NMIJ\mathchar`-F2}}u_{\mathrm{B}}^{\mathrm{NMIJ\mathchar`-F2}}u_{\mathrm{B}}^{\mathrm{NMIJ\mathchar`-F2\,}m_{2}}\nonumber\\
&&+w_{\mathrm{August21}}^{\mathrm{NMIJ\mathchar`-F2}}\sqrt{\frac{T_{m_{1},m_{2}}}{T_{m_{2}}}}(u_{\mathrm{A}}^{\mathrm{NMIJ\mathchar`-F2}}u_{\mathrm{A}}^{\mathrm{NMIJ\mathchar`-F2\,}m_{2}}+u_{\mathrm{cable}}^{\mathrm{NMIJ\mathchar`-F2}}u_{\mathrm{cable}}^{\mathrm{NMIJ\mathchar`-F2\,}m_{2}})],\quad\quad
\label{correlationeq}
\end{eqnarray}
where $u_{\mathrm{B}}^{\mathrm{Yb\,}m_{2}}(=1.4\times10^{-16}\sim u_{\mathrm{B}}^{\mathrm{Yb}})$ and $u_{\mathrm{B}}^{\mathrm{NMIJ\mathchar`-F2\,}m_{2}}(=4.7\times10^{-16}\sim u_{\mathrm{A}}^{\mathrm{NMIJ\mathchar`-F2}})$ denote the systematic uncertainties of NMIJ-Yb1 and NMIJ-F2 in $m_{2}$, respectively, $w_{\mathrm{total}}^{\mathrm{NMIJ\mathchar`-F2}}(=0.033)$ the total weight involving NMIJ-F2 for $m_{1}$ described in Fig.~\ref{eachpfsfig}, and $w_{\mathrm{August21}}^{\mathrm{NMIJ\mathchar`-F2}}(=0.004)$ the specific weight in August 2021 for $m_{1}$. $u_{\mathrm{A}}^{\mathrm{NMIJ\mathchar`-F2\,}m_{2}}(=1.2\times10^{-16})$ and $u_{\mathrm{cable}}^{\mathrm{NMIJ\mathchar`-F2\,}m_{2}}(=1.5\times10^{-16})$ are the statistical uncertainty of NMIJ-F2 and the link uncertainty evaluated for NMIJ-F2 in $m_{2}$, respectively. The contribution from $u_{\mathrm{A}}^{\mathrm{Yb}}$ is neglected in Eq.~(\ref{correlationeq}), since the statistical uncertainty is dominated by NMIJ-F2.

\begin{table}[t]
\caption{Correlation coefficients between measurements involving NMIJ-Yb1.}  
	\label{corelationresults}
	\begin{center} 
\begin{tabular}{ll}
\hline
Notation  & Estimated value \\
\hline
$r_{m_{1},m_{2}}$ & 0.319\\
$r_{m_{1},m_{3}}$ & 0.250\\
$r_{m_{1},m_{4}}$ & 0.099\\
$r_{m_{2},m_{3}}$ & 0.082\\
$r_{m_{2},m_{4}}$ & 0.054\\
$r_{m_{3},m_{4}}$ & 0.748 \cite{Margolis2024}\\
\hline
\end{tabular}
\end{center}
\end{table}

In the estimation of the correlation coefficient $r_{m_{1},m_{3}}$ between the present measurement $m_{1}$ and the previous 6-month measurement $m_{3}$ using the TAI link, we take account of the fact that these two measurements share a common portion $u_{\mathrm{B\,com}}^{\mathrm{Yb}\,m_{1},m_{3}}=8.8\times10^{-17}$ of $u_{\mathrm{B}}^{\mathrm{Yb}}$, $u_{\mathrm{B/Lab}}^{\mathrm{Yb}}$, and the systematic uncertainties of NIM5, SYRTE-FO1, SYRTE-FO2, SYRTE-FOM, PTB-CSF1, PTB-CSF2, and SU-CsFO2. The statistical uncertainties are not correlated, since there are no overlapping periods. The correlation coefficient is given by 
\begin{eqnarray}
r_{m_{1},m_{3}}&=&\frac{1}{u_{\mathrm{total}}^{m_{1}}u_{\mathrm{total}}^{m_{3}}}[(u_{\mathrm{B\,com}}^{\mathrm{Yb}\,m_{1},m_{3}})^{2}+(u_{\mathrm{B/Lab}}^{\mathrm{Yb}})^{2}\nonumber\\
&&+\sum_{s}\sum_{p}w_{p}^{s\,m_{1}}u_{\mathrm{B}\,p}^{s\,m_{1}}\sum_{q}w_{q}^{s\,m_{3}}u_{\mathrm{B}\,q}^{s\,m_{3}}],
\end{eqnarray}
where $w_{p}^{s\,m_{1}}$ ($u_{\mathrm{B}\,p}^{s\,m_{1}}$) and $w_{q}^{s\,m_{3}}$ ($u_{\mathrm{B}\,q}^{s\,m_{3}}$) denote the weights (systematic uncertainties) of a PFS with index $s$ at month $p$ in $m_{1}$ and of a PFS with index $s$ at month $q$ in $m_{3}$, respectively. The summation over $s$ is carried out for the common 7 PFSs. While $w_{p}^{s\,m_{1}}$ is simply taken from the obtained $\boldsymbol{w}$ in the present analysis, $w_{q}^{s\,m_{3}}$ is derived in a different way. In $m_{3}$, we employed a frequency $y(\mathrm{TAI-PSFS})$ of TAI referenced an ensemble of PSFS averaged over a month, which is calculated by BIPM \cite{circulart}. From the data in Circular T, a weight $w_{q}^{s\,q}$ of an individual PFS for $y(\mathrm{TAI-PSFS})$ at a specific month $q$ is estimated \cite{Hachisu2016}. The weight $w_{q}^{s\,m_{3}}$ for the 6-month measurement $m_{3}$ is then calculated by multiplying $w_{q}^{s\,q}$ by a monthly weight $w_{q}$ for an overall mean $\overline{y(\mathrm{Yb-PSFS})}$ in $m_{3}$, which is derived in the analysis of $m_{3}$ (Fig.~8 of Ref.~\cite{Kobayashi2020}). 

The other correlation coefficients $r_{m_{1},m_{4}}$, $r_{m_{2},m_{3}}$, and $r_{m_{2},m_{4}}$ are small compared with $r_{m_{1},m_{2}}$ and $r_{m_{1},m_{3}}$, because of the reevaluation of the systematic uncertainty of NMIJ-Yb1 \cite{Kobayashi2022}, no correlations arising from PFS, and no overlapping periods. We estimate these coefficients as follows,
\begin{equation}
r_{m_{1},m_{4}}=\frac{(u_{\mathrm{B\,com}}^{\mathrm{Yb}\,m_{1},m_{4}})^{2}}{u_{\mathrm{total}}^{m_{1}}u_{\mathrm{total}}^{m_{4}}},
\end{equation}

\begin{equation}
r_{m_{2},m_{3}}=\frac{(u_{\mathrm{B\,com}}^{\mathrm{Yb}\,m_{2},m_{3}})^{2}+(u_{\mathrm{B/Lab}}^{\mathrm{Yb}})^{2}}{u_{\mathrm{total}}^{m_{2}}u_{\mathrm{total}}^{m_{3}}},
\end{equation}

\begin{equation}
r_{m_{2},m_{4}}=\frac{(u_{\mathrm{B\,com}}^{\mathrm{Yb\,}m_{2},m_{4}})^{2}}{u_{\mathrm{total}}^{m_{2}}u_{\mathrm{total}}^{m_{4}}},
\end{equation}
where $u_{\mathrm{B\,com}}^{\mathrm{Yb}\,m_{i},m_{j}}$ denotes the common systematic uncertainty of NMIJ-Yb1 between $m_{i}$ and $m_{j}$. The common uncertainties are estimate as $u_{\mathrm{B\,com}}^{\mathrm{Yb}\,m_{1},m_{4}}=u_{\mathrm{B\,com}}^{\mathrm{Yb}\,m_{1},m_{3}}$ and $u_{\mathrm{B\,com}}^{\mathrm{Yb}\,m_{2},m_{3}}=u_{\mathrm{B\,com}}^{\mathrm{Yb}\,m_{2},m_{4}}=1.1\times10^{-16}$.

\section{Conclusion}
In conclusion, we report the absolute frequency measurement of $^{171}$Yb with an uncertainty of $1.9\times10^{-16}$ by comparing NMIJ-Yb1 and 13 PFSs via TAI. We also derive the correlation coefficients between the present measurement and our previous measurements, which is important for updating the recommended frequency. This work is one of the essential contributions towards the redefinition of the SI second. 

\section*{Acknowledgments}
We are grateful to F.-L. Hong and Y. Hisai for discussions and assistances to develop NMIJ-Yb1. We thank A. Iwasa and Y. Fujii for the support to maintain UTC(NMIJ). We appreciate A. Takamizawa for operating NMIJ-F2 and discussions about clock comparisons. We acknowledge Geospatial Information Authority of Japan for the determination of the geopotential value of NMIJ-Yb1. We thank M. T\o nnes and N. Nemitz for giving us the insightful presentation about the analysis with covariance matrices in the OptAsia collaboration meeting. We are indebted to national metrology institutes for operating the primary and secondary frequency standards and the BIPM time department for making the evaluation results available in Circular T. This work was supported by Japan Society for the Promotion of Science (JSPS) KAKENHI Grant Number 17H01151, 17K14367, and 22H01241, JST-Mirai Program Grant Number JPMJMI18A1, and the JST Moonshot R $\&$ D Program Grant Number JPMJMS2268, Japan.

\end{document}